\documentclass[12pt]{article}
\usepackage{amsmath}
\usepackage{latexsym}
\usepackage{epsfig}
\usepackage{graphics}
\title{{\bf Analytical On-shell Calculation of Low Energy Higher Order Scattering}}
\author{Barry R. Holstein\\
Department of Physics-LGRT\\
University of Massachusetts\\
Amherst, MA  01003\\
and\\
Kavli Institute for Theoretical Physics\\
University of California\\
Santa Barbara, CA  93016}
\begin{titlepage}
\begin{document}
\maketitle
\begin{abstract}
We demonstrate that the use of analytical on-shell methods involving calculation of the discontinuity across the t-channel cut associated with the exchange of a pair of massless particles (photons or gravitons) can be used to evaluate one-loop contributions to electromagnetic and gravitational scattering, with and without polarizability, reproducing via simple algebraic manipulations, results obtained previously, generally using Feynman diagram techniques.  In the gravitational case the use of factorization permits a straightforward and algebraic calculation of higher order scattering without consideration of ghost contributions or of triple-graviton couplings, which made previous evaluations considerably more arduous.
\end{abstract}
\end{titlepage}

\section{Introduction}

The calculation of scattering amplitudes is a staple of theoretical physics, and recently a number of investigations have been reported which study higher order effects in
electromagnetic scattering~\cite{Iwa71},\cite{Spr93},\cite{Fei88},\cite{Ros08}, gravitational scattering~\cite{Don94},\cite{Muz95},\cite{Ham95},\cite{Akh97},\cite{Khr03},\cite{Bje03}\cite{Ros08a} and both~\cite{But06},\cite{Fal08},\cite{Ros08b}.  The goal of such calculations has typically been to find an effective potential which characterizes these higher order effects.  For both electromagnetic and gravitational interactions, the leading potential is, of course, well-known and has the familiar $1/r$ fall-off with distance.  The higher order contributions required by quantum mechanics lead to corrections which are shorter range, from local to polynomial fall-off as $1/r^n$ with $n\ge 2$.  This effective potential is defined to be the Fourier transform of the nonrelativistic scattering amplitude via\footnote{Note that Eq. (\ref{eq:nb}) follows from the Born approximation for the scattering amplitude
\begin{equation}
{\rm Amp}(\boldsymbol{q})=<\boldsymbol{p}_f|\hat{V}|\boldsymbol{p}_i>=\int d^3re^{i\boldsymbol{q}\cdot\boldsymbol{r}}V(r)
\end{equation}
and nonrelativistic amplitudes are defined by taking the low energy limit and dividing the covariant forms by the normalizing factor $4E_AE_B\simeq 4m_Am_B$.}
\begin{equation}
V(r)=-\int{d^3q\over (2\pi)^3}e^{-i\boldsymbol{q}\cdot\boldsymbol{r}}{\rm Amp}(\boldsymbol{q})\label{eq:nb}
\end{equation}
where $\boldsymbol{q}=\boldsymbol{p}_i-\boldsymbol{p}_f$ is the three-momentum transfer.  Then, for lowest order one-photon or one-graviton exchange, the dominant momentum-transfer dependence arises from the massless propagator ${1\over q^2}\stackrel{NR}{\longrightarrow}{-1\over\boldsymbol{q}^2}+{\cal O}(\boldsymbol{q}^4)$, whose Fourier transform
\begin{equation}
\int{d^3q\over (2\pi)^3}e^{-i\boldsymbol{q}\cdot\boldsymbol{r}}{1\over \boldsymbol{q}^2}={1\over 4\pi|\boldsymbol{r}|}
\end{equation}
yields the well-known $1/r$ dependence.  By dimensional analysis, it is clear that shorter-range $1/r^2$ and $1/r^3$ behavior can arise only from {\it nonanalytic} $1/|\boldsymbol{q}|$ and $\ln \boldsymbol{q}^2$ dependence, which are associated with higher order scattering contributions.  Analytic momentum dependence from polynomial contributions in such diagrams leads only to short-distance ($\delta^3(\boldsymbol{r})$ and its derivatives) effects.  Thus, if we are seeking the long-range corrections, we need identify only the nonanalytic components of the higher order low energy contributions to the scattering amplitude.

The basic idea behind use of on-shell methods is that the scattering amplitude must satisfy the stricture of unitarity, which requires that
its discontinuity across the right hand cut is given by
\begin{equation}
 {\rm Disc}\, T_{fi}=i(T_{fi}-{T^\dagger}_{fi})=-\sum_n T_{fn}T^\dagger_{ni}\label{eq:mk}
\end{equation}
By requiring that Eq. ({\ref{eq:mk}) be satisfied, we guarantee that the correct nonanalytic structure will be maintained, and below we demonstrate how this program can be carried out in the case of the electromagnetic and gravitational scattering of spinless particles, with and without polarizability effects.\footnote{It is interesting to note that this method is essentially the one used by Feynman in his seminal paper, wherein he quantized gravity and realized the need for the introduction of ghosts~\cite{Fey63}.}  We will show how results obtained previously using Feynman diagram techniques can be obtained by {\it much} simplified analytical on-shell methods.  This simplification arises essentially due the interchange of the order of integration and summation.  That is, in the conventional Feynman technique, one evaluates separate (four-dimensional) Feynman integrals for each diagram, which are then summed.  In the on-shell method, one first sums over the Compton scattering diagrams to obtain helicity amplitudes and {\it then} performs a (two-dimensional) solid-angle integration. There are a number of reasons why the latter procedure is more efficient.  For one, by using the explicitly {\it gauge-invariant} Compton and gravitational Compton amplitudes, the decomposition into separate {\it gauge-dependent} diagrams is avoided.  Secondly, the various statistical/combinatorial factors are included automatically. Thirdly, because the intermediate states are on-shell, there is no gravitational ghost contribution~\cite{Pes95}.  Finally, the evaluation of gravitational Compton amplitudes allows the use of factorization, which ameliorates the need to include the triple graviton coupling associated with the graviton pole diagram~\cite{Cho95},\cite{Ber02}.  The superposition of all these effects allows a relatively simple and highly efficient algebraic calculation of both the electromagnetic and gravitational scattering amplitudes.  (One indication of the simplicity afforded by this method, in the gravitational case, is that between the seminal 1994 work of Donoghue~\cite{Don94} and the 2003 papers by \cite{Khr03} and \cite{Bje03}, there were a number of reported Feynman diagram calculations of gravitational scattering which contained errors~\cite{Muz95},\cite{Ham95},\cite{Akh97}.)

\section{Electromagnetic Scattering}

We begin with the case of electromagnetic scattering of spinless particles of mass $m_A$ and $m_B$ respectively.   The  $t$-channel Compton amplitude ({\it i.e.,} the amplitude for two spinless particles of charge $e$ and mass $m_A$ to annihilate into a pair of photons) is well known~\cite{Hol14}
\begin{equation}
^{A}{\rm Amp}_0^{em}=2e^2\left(\epsilon_1^*\cdot\epsilon_2^*-{\epsilon_1^*\cdot p_1\epsilon_2^*\cdot p_2\over p_1\cdot k_1}-{\epsilon_1^*\cdot p_2\epsilon_2^*\cdot p_1\over p_1\cdot k_2}\right)
\end{equation}
It is convenient to use the helicity formalism~\cite{Jac59}, where helicity is defined as the projection of the photon spin on its momentum axis.  The helicity amplitudes for $t$-channel spin-0-spin-0 Compton annihilation are found, in the center of mass frame, to have the form~\cite{Hol06}
\begin{eqnarray}
^AA_0^{EM}(++)&=&^AA_0^{EM}(--)=2e^2\left({{m_A^2\over E_A^2-\boldsymbol{p}_A^2\cos^2\theta_A}}\right)\,,\nonumber\\
^AA_0^{EM}(+-)&=&^AA_0^{EM}(-+)=2e^2\left({\boldsymbol{p}_A^2\sin^2\theta_A\over E_A^2-\boldsymbol{p}_A^2\cos^2\theta_A}\right)\,,
\end{eqnarray}
where $m_A,\,E_A,\,\pm\boldsymbol{p}_A$ are the mass, energy, momentum of the spinless particles and $\theta_A$ the angle of the outgoing photon with respect to the incoming target particle---$\cos\theta_A=\hat{\boldsymbol{p}}_A\cdot\hat{\boldsymbol{k}}$.   It was shown by Feinberg and Sucher that the annihilation amplitudes $A+A'\rightarrow\gamma_1+\gamma_2$ and $\gamma_1+\gamma_2\rightarrow B+B'$ needed in the unitarity relation, Eq. (\ref{eq:mk}), can be generated by making an analytic continuation to imaginary momentum $\boldsymbol{p}_i\rightarrow im_i\xi_i\hat{\boldsymbol{p}}_i$, where $\xi_i^2=1-{t\over 4m_i^2}$ with $i=A,\,B$ and $t=(p_A+p_A')^2$ is the t-channel Mandelstam variable~\cite{Fei88}.  Then
\begin{eqnarray}
^iA^{em}_0(++)&=&^iA^{em}_0(--)= 2e^2{1+\tau_i^2\over d_i}\,,\nonumber\\
^iA^{em}_0(+-)&=&^iA^{em}_0(-+)= 2e^2{1-x_i^2\over d_i}\,,\label{eq:bv}
\end{eqnarray}
where we have defined $\tau_i=\sqrt{t}/2m_i\xi_i$, $x_i=\hat{\boldsymbol{p}}_i\cdot\hat{\boldsymbol{k}_1}$, and $d_i=\tau_i^2+x_i^2$.
Equivalently Eq. (\ref{eq:bv}) can be represented succinctly via
\begin{equation}
^iA^{em}_0(ab)=2e^2{\cal O}_i^{jk}{\epsilon_{1j}^{a*}}{\epsilon_{2k}^{b*}}\end{equation}
where
\begin{equation}
{\cal O}_i^{jk}={1\over d_i}\left(d_i\delta^{jk}+2\hat{p}_i^j\hat{p}_i^k\right)\quad i=A,\,B
\end{equation}
Substituting in Eq. (\ref{eq:mk}), we determine the discontinuity of the scattering amplitude of spinless particles having masses $m_A,m_B$ across the $t$-channel two-photon cut in the CM frame\footnote{Note that we have divided the scattering amplitude by the normalizing factor $4m_Am_B$ since it will be used in the nonrelativistic limit.}
\begin{eqnarray}
&&{\rm Disc}\,{\rm Amp}_2^{em}(q)=-{i\over 2!}{(2e^2)^2\over 4m_Am_B}\int {d^3k_1\over (2\pi)^32k_{10}}{d^3k_2\over (2\pi)^32k_{20}}(2\pi)^4\delta^4(p_1+p_2-k_1-k_2)\nonumber\\
&\times&\sum_{a,b=1}^2\left[{\cal O}_A^{ij}\epsilon^{a*}_{1i}\epsilon^{b*}_{2j}\epsilon^a_{1k}\epsilon^b_{2\ell}{\cal O}_B^{k\ell*}\right]=-i{e^4\over 16\pi m_Am_B}<\sum_{i,j,k,\ell=1}^3{\cal O}_A^{ij}\delta^T_{ik}\delta^T_{j\ell}{\cal O}_B^{k\ell*}>\nonumber\\
\quad
\end{eqnarray}
where
\begin{equation}
\delta^T_{ik}=\sum_{a=1}^2\epsilon_i^{a*}\epsilon_k^{a}=\delta_{ik}-\hat{k}_i\hat{k}_k
\end{equation}
represents the sum over photon polarizations and
$$<G>\equiv \int{d\Omega_{\hat\,{\boldsymbol{k}}}\over 4\pi}G$$
defines the solid-angle average.  Performing the indicated polarization sums, we find
\begin{eqnarray}
&&<\sum_{i,j,k,\ell=1}^3{\cal O}_A^{ij}\delta^T_{ik}\delta^T_{j\ell}{\cal O}_B^{k\ell*}>=<{1\over d_Ad_B}\left(4(y-x_Ax_B)^2-2(1-x_A^2)(1-x_B^2)\right.\nonumber\\
&+&\left.2(1+\tau_A^2)(1+\tau_B^2)\right)>\stackrel{t<<m_A^2,m_B^2}{\longrightarrow}<{1\over d_Ad_B}\left(4(y-x_Ax_B)^2+2x_A^2+2x_B^2-2x_A^2x_B^2\right)>\nonumber\\
\quad
\end{eqnarray}
where
\begin{equation}
y(s,t)=\hat{\boldsymbol{p}}_A\cdot\hat{\boldsymbol{p}}_B={2s+t-2m_A^2-2m_B^2\over 4m_A\xi_Am_B\xi_B}\stackrel{s\rightarrow s_0}{\longrightarrow} 1+{\cal O}(t)\,.
\end{equation}
characterizes the angle between incoming and outgoing spinless particles.  Near threshold---$s_0=(m_A+m_B)^2$---$y(s_0,t)=1+{\cal O}(t)$ and
\begin{eqnarray}
&&{\rm Disc}\,{\rm Amp}_2^{em}(q)\simeq -i{e^4\over 8\pi m_Am_B}<{1\over d_Ad_B}\left(2y^2-4yx_Ax_B+x_A^2+x_B^2+x_A^2x_B^2\right)>\nonumber\\
&&\stackrel{y\rightarrow 1}{\longrightarrow}-i{e^4\over 8\pi m_Am_B}<{1\over d_Ad_B}\left(2-4x_Ax_B+x_A^2+x_B^2+x_A^2x_B^2\right)>\label{eq:gb}
\end{eqnarray}
Defining
\begin{equation}
I_{nm}\equiv <{x_A^nx_B^m\over d_Ad_B}>\,,
\end{equation}
the required angular integrals $I_{00},\,I_{11}$ have been given by Feinberg and Sucher~\cite{Fei70}, and higher order forms can be found by use of the identities $x_i^2=d_i-\tau_i^2,\,i=A,B$, yielding
\begin{eqnarray}
{\rm Disc}\,{\rm Amp}_2^{em}(q)&\simeq&-i{e^4\over 8\pi m_Am_B}\left[2I_{00}-4I_{11}+I_{20}+I_{02}+I_{22}\right]\nonumber\\
&=&-i{e^4\over 8\pi m_Am_B}\left[{\pi(m_A+m_B)\over \sqrt{t}}+{7\over 3}+i{4\pi m_Am_Bm_r\over p_0t}+\ldots\right]\nonumber\\
\quad
\end{eqnarray}
where $p_0=\sqrt{{m_r(s-s_0)\over m_A+m_B}}$ is the center of mass momentum for the spinless scattering process and $m_r=m_Am_B/(m_A+m_B)$ is the reduced mass. Since
\begin{equation}
{\rm Disc}\,\left[\log(-t),\,\sqrt{1\over -t}\right]=\left[2\pi i,\,-i{2\pi^2\over \sqrt{t}}\right]
\end{equation}
the scattering amplitude is
\begin{equation}
{\rm Amp}_2^{em}(q)=-{e^4\over 16\pi^2 m_Am_B}\left[{7\over 3}L-S(m_A+m_B)+4\pi i{m_rm_Am_B\over p_0t}L+\ldots\right]\,,\label{eq:vc}
\end{equation}
where we have defined $L=\log(-t)$ and $S=\pi^2/\sqrt{-t}$.  The imaginary component of Eq. (\ref{eq:vc}) represents the Coulomb phase, or equivalently the contribution of the second Born approximation, which must be subtracted in order to define a proper higher-order potential.  Using~\cite{Dal51}
\begin{eqnarray}
B_2^{em}(q)&=&i\int{d^3\ell\over (2\pi)^3}{e^2\over |\boldsymbol{p}_f-\boldsymbol{\ell}|^2+\lambda^2}
{i\over {p_0^2\over 2m_r}-{\ell^2\over 2m_r}+i\epsilon}{e^2\over |\boldsymbol{\ell}-\boldsymbol{p}_i|^2+\lambda^2}\nonumber\\
&=&-i{e^4\over 4\pi}{m_r\over p_0}{\log(-t)\over t}\label{eq:cx}
\end{eqnarray}
what remains is the higher order electromagnetic amplitude we are seeking, leading to the effective potential
\begin{eqnarray}
V_2^{em}(r)&=&-\int{d^3q\over (2\pi)^3}e^{-i\boldsymbol{q}\cdot\boldsymbol{r}}\left({\rm Amp}_2^{em}(q)-B_2^{em}(q)\right)\nonumber\\
&=&-{\alpha_{em}^2(m_A+m_B)\over 2r^2}-{7\alpha_{em}^2\hbar\over 6\pi r^3}\label{eq:dz}
\end{eqnarray}
where $\alpha_{em}={e^2\over 4\pi}$ is the fine structure constant, in agreement with the form calculated by Feynman diagram methods~\cite{Ros08}.

One can also calculate the interaction of a charged and a neutral polarizable spinless system.  For the neutral system we use the Hamiltonian
\begin{eqnarray}
H_{eff}&=&-{1\over 2}(4\pi\alpha_E\boldsymbol{E}^2+4\pi\beta_M\boldsymbol{H}^2)\nonumber\\
&=&-{4\pi\alpha_E\over 2m_A^2}p_1^\alpha F_{\alpha\beta}F^{\beta\gamma}p_{1\gamma}-{4\pi\beta_M\over 8m_A^2}\epsilon^{\alpha\beta\gamma\delta}
{\epsilon_\alpha}^{\rho\sigma\lambda}F_{\rho\sigma}p_{1\lambda}
\end{eqnarray}
where $\alpha_E,\,\beta_M$ are the electric,\,magnetic dipole polarizabilities and $F_{\mu\nu}$ is the electromagnetic field tensor, which yields the contact helicity amplitude for emission of a photon pair
\begin{equation}
N_A^{ab}=\pi t{\cal V}_A^{ij}\epsilon^{a*}_{1i}\epsilon^{b*}_{2j}
\end{equation}
with
\begin{equation}
{\cal V}_A^{ij}=(\alpha_E^Ax_A^2-\beta_M^A(2-x_A^2))\delta^{ij}+2(\alpha_E^A+\beta_M^A)\hat{p}_A^i\hat{p}_A^j
\end{equation}
Using the angular averaged integral $J^B_{00}=<{1\over d_B}>=({\pi\over 2}-{\rm tan}^{-1}\tau_B)/\tau_B$~\cite{Fei88} and higher order forms $J^B_{nm}=<{x_A^nx_B^m\over d_B}>$ generated via the use of the substitutions $x_i^2=d_i-\tau_i^2,\,i=A,B$, the discontinuity for spinless charged-neutral scattering is
\begin{eqnarray}
&&{\rm Disc}\,{\rm Amp}_2^{N-em}=-{i\over 2!}\int {d^3k_1\over (2\pi)^32k_1^0}{d^3k_2\over (2\pi)^32k_2^0}(2\pi)^4\delta^4(p_1+p_2-k_1-k_2)\nonumber\\
&\times&\sum_{a,b=1}^2{\pi e^2t\over m_B}{\cal V}_A^{ij}\epsilon^{a*}_{1i}\epsilon^{b*}_{2j}\epsilon^a_{1k}\epsilon^b_{2\ell}{\cal O}_B^{k\ell*}={-ie^2t\over 16m_B}
\sum_{i,j,k,\ell=1}^3<\left[{\cal V}_A^{ij}P^T_{ik}P^T_{j\ell}{\cal O}_B^{k\ell*}\right]>\nonumber\\
&\stackrel{y\rightarrow 1}{\longrightarrow}& -2\pi i{\alpha_{em} t\over 4m_B}\left[2\alpha_E^AJ^B_{00}+(\alpha_E^A+\beta_M^A)\left(J^B_{02}+J^B_{20}-4J^B_{11}+J^B_{22}\right)\right]\nonumber\\
&=& -2\pi i{\alpha_{em} t\over 4m_B}\left[({2\pi m_B\over \sqrt{-t}}-{11\over 3})\alpha_E^A-{5\over 3}\beta_M^A\right]
\end{eqnarray}
The scattering amplitude is then
\begin{equation}
{\rm Amp}_2^{N-em}(q)=\alpha_{em}\alpha_E^A{\pi^2\sqrt{-t}\over 2}+{1\over 3}\alpha_{em}(11\alpha_E^A+5\beta_M^A)t\log -t
\end{equation}
and the effective potential describing interaction of a charged and neutral system
\begin{equation}
V_2^{N-em}(r)=-\int{d^3q\over (2\pi)^3}e^{-i\boldsymbol{q}\cdot\boldsymbol{r}}{\rm Amp}_2^{N-em}(q)=-{\alpha_{em}\alpha_E^A\over 2r^4}+\alpha_{em}{(11\alpha_E^A+5\beta_M^A)\hbar\over 4\pi m_B r^5}
\end{equation}
in agreement with well-known forms~\cite{Ber76},\cite{Fei92}.

Finally, we can examine the interaction of two spinless systems, both of which are characterized by polarizabilities~\cite{Hol14a}.  Defining the angular averaged quantities $K_{nm}=<x_A^nx_B^m>$, we have
\begin{eqnarray}
&&{\rm Disc}\,{\rm Amp}_2^{NN-em}(q)=-{i\over 2!}\int {d^3k_1\over (2\pi)^32k_1^0}{d^3k_2\over (2\pi)^32k_2^0}(2\pi)^4\delta^4(p_1+p_2-k_1-k_2)\nonumber\\
&\times&\sum_{a,b=1}^2\pi^2t^2{\cal V}_A^{ij}\epsilon^{a*}_{1i}\epsilon^{b*}_{2j}\epsilon^a_{1k}\epsilon^b_{2\ell}{\cal V}_B^{k\ell*}=-i{\pi t^2\over 16}\sum_{i,j,k,\ell=1}^3<\left[{\cal V}_A^{ij}P^T_{ik}P^T_{j\ell}{\cal V}_B^{k\ell*}\right]>\nonumber\\
&=&-i2\pi{t^2\over 16}\left[(\alpha_E^A\alpha_E^B+\beta_M^A\beta_M^B)(2K_{00}+K_{20}+K_{02}-4K_{11}+K_{22})\right.\nonumber\\
&+&\left.(\alpha_E^A\beta_M^B+\alpha_E^B\beta_M^A)(K_{20}+K_{02}-4K_{11}+K_{22})\right]\nonumber\\
&=&-i2\pi{t^2\over 16}\left[(\alpha_E^A\alpha_E^B+\beta_M^A\beta_M^B){23\over 15}-(\alpha_E^A\beta_M^B+\alpha_E^B\beta_M^A){7\over 15}\right]\nonumber\\
\quad
\end{eqnarray}
yielding the scattering amplitude
\begin{equation}
{\rm Amp}_2^{NN-em}(q)={t^2L\over 16}\left[(\alpha_E^A\alpha_E^B+\beta_M^A\beta_M^B){23\over 15}-(\alpha_E^A\beta_M^B+\alpha_E^B\beta_M^A){7\over 15}\right]
\end{equation}
and the effective potential
\begin{eqnarray}
V_2^{NN-em}(r)&=&-\int{d^3q\over (2\pi)^3}e^{-i\boldsymbol{q}\cdot\boldsymbol{r}}{\rm Amp}_2^{NN-em}(q)\nonumber\\
&=&{-23(\alpha_E^A\alpha_E^B+\beta_M^A\beta_M^B)+7(\alpha_E^A\beta_M^B+\alpha_E^B\beta_M^A)\over 4\pi r^7}\,,\label{eq:rx}
\end{eqnarray}
which has the familiar Casimir-Polder form~\cite{Cas48},\cite{Hol01}.

\section{Gravitational Scattering}

These electromagnetic results were first obtained using related methods by Feinberg and Sucher~\cite{Fei88},\cite{Fei92}.  The real power and simplicity of the on-shell methods, however, is found in the {\it gravitational} scattering case, where what is needed is the amplitude for the annihilation of a spinless pair of particles of mass $m_A$ and $m_B$ connected by a two-{\it graviton} intermediate state. A major simplification in this regard is provided by factorization, which avers that the gravitational Compton scattering amplitude is given in terms of the product of ordinary Compton scattering amplitudes multiplied by a simple kinematic factor~\cite{Cho95},\cite{Ber02}.  That is, defining
\begin{equation}
F={p_i\cdot k_ip_i\cdot k_f\over k_i\cdot k_f}\,,
\end{equation}
we have the remarkable identity for the gravitational Compton amplitude $^A{\rm Amp}_0^{grav}$
\begin{eqnarray}
^A{\rm Amp}_0^{grav}&=&{\kappa^2\over 8e^4}F\left(^A{\rm Amp}_0^{em}\right)^2\nonumber\\
&=&{\kappa^2\over 2}\left({p_i\cdot k_ip_f\cdot k_f\over k_i\cdot k_f}\right)\left({\epsilon_i\cdot p_i\epsilon_f^*\cdot p_f\over p_i\cdot k_i}-{\epsilon_i\cdot p_f\epsilon_f^*\cdot p_i\over p_i\cdot k_f}-\epsilon_i\cdot\epsilon_f^*\right)\nonumber\\
&\times&\left({\epsilon_i\cdot p_i\epsilon_f^*\cdot p_f\over p_i\cdot k_i}-{\epsilon_i\cdot p_f\epsilon_f^*\cdot p_i\over p_i\cdot k_f}-\epsilon_i\cdot\epsilon_f^*\right)\label{eq:nx}
\end{eqnarray}
Using Eq. (\ref{eq:nx}), the two-graviton helicity amplitudes can be written in the factorized form
\begin{eqnarray}
^AB_0^{grav}(ij)&=&{\kappa^2m_A^2\xi_A^2d_A\over 4}{\cal O}_A^{rs}\epsilon^{*i}_r\epsilon^{*j}_s{\cal O}_A^{uv}\epsilon^{*i}_u\epsilon^{*j}_v\nonumber\\
^AB_0^{grav}(ij)&=&{\kappa^2m_B^2\xi_B^2d_B\over 4}\epsilon^{i}_r\epsilon^{j}_s{\cal O}_B^{rs}\epsilon^{i}_u\epsilon^{j}_v{\cal O}_B^{uv}
\end{eqnarray}
We can then write the unitarity relation for the gravitational scattering of spinless particles of mass $m_A,\,m_B$ as
\begin{eqnarray}
&&{\rm Disc}\,{\rm Amp}_2^{grav}(q)=-{i\over 2!}{\kappa^4m_A^2\xi_A^2m_B^2\xi_B^2\over 64m_Am_B}\int {d^3k_1\over (2\pi)^32k_{10}}{d^3k_2\over (2\pi)^32k_{20}}\nonumber\\
&\times&(2\pi)^4\delta^4(p_1+p_2-k_1-k_2)d_Ad_B\sum_{r,s=1}^2\left[{\cal O}_A^{ij}{\cal O}_A^{k\ell}\epsilon^{r*}_{1i}\epsilon^{s*}_{2j}\epsilon^{r*}_{1k}\epsilon^{s*}_{2\ell}\epsilon^r_{1a}\epsilon^s_{2b}\epsilon^r_{1c}\epsilon^s_{2d}{\cal O}_B^{ab*}{\cal O}_B^{cd*}\right]\nonumber\\
&=&-i{\kappa^4m_A^2\xi_A^2m_B^2\xi_B^2\over 1024\pi m_Am_B}<d_Ad_B\sum_{i,j,k,\ell=1}^3\sum_{a,b,c,d=1}^3{\cal O}_A^{ij}{\cal O}_A^{k\ell}P^G_{ik;ac}P^G_{j\ell;bd}{\cal O}_B^{ab*}{\cal O}_B^{cd*}>\nonumber\\
\quad
\end{eqnarray}
where the sum over graviton polarizations is
\begin{equation}
P^G_{ik;ac}=\sum_{r=1}^2\epsilon^{r*}_i\epsilon^{r*}_k\epsilon^{r}_a\epsilon^{r}_c={1\over 2}\left[\delta^T_{ia}\delta^T_{kc}+\delta^T_{ic}\delta^T_{ka}-\delta^T_{ik}\delta^T_{ac}\right]
\end{equation}
Performing the polarization sum, we find
\begin{eqnarray}
&&<\left[d_Ad_B{\cal O}_A^{ij}{\cal O}_A^{k\ell}P^G_{ik;ac}P^G_{j\ell;bd}{\cal O}_B^{ab*}{\cal O}_B^{cd*}\right]>=
<{1\over d_Ad_B}\left[4\left(2(y-x_Ax_B)^2\right.\right.\nonumber\\
&-&\left.\left.(1-x_A^2)(1-x_B^2)\right)^2-2(1-x_A^2)^2(1-x_B^2)^2+2(1+\tau_A^2)^2(1+\tau_B^2)^2\right]>\nonumber\\
&\stackrel{t<<m_A^2,m_B^2}{\longrightarrow}&<{2\over d_Ad_B}\left[
8(y-x_Ax_B)^4\right.\nonumber\\
&-&\left.8(y-x_Ax_B)^2(1-x_A^2)(1-x_B)^2+(1-x_A^2)^2(1-x_B^2)^2+1\right]>
\end{eqnarray}
Using the angular-averaged forms $I_{nm}$ defined earlier, we determine then
\begin{eqnarray}
&&{\rm Disc}\,{\rm Amp}^{grav}_2(q)\simeq-i{\kappa^4m_A\xi_A^2m_B\xi_B^2\over 512\pi}<{1\over d_Ad_B}\left[
8(y-x_Ax_B)^4\right.\nonumber\\
&-&\left.8(y-x_Ax_B)^2(1-x_A^2)(1-x_B)^2+(1-x_A^2)^2(1-x_B^2)^2+1\right]>\nonumber\\
&=&-i{\kappa^4m_A\xi_A^2m_B\xi_B^2\over 512\pi}\left[I_{44}+6I_{42}+6I_{24}-16I_{33}-16I_{13}-16I_{31}\right.\nonumber\\
&+&\left.I_{40}+I_{04}+36I_{22}+6I_{20}+6I_{02}-16I_{11}+2I_{00}\right]\nonumber\\
&=&-i{\kappa^4m_Am_B\over 512\pi}\left[{41\over 5}+6{\pi(m_A+m_B)\over \sqrt{t}}+4\pi i{m_Am_Bm_r\over p_0t}+\ldots\right]\label{eq:mg}
\end{eqnarray}
so that the gravitational scattering amplitude is
\begin{equation}
{\rm Amp}_2^{grav}(q)=-{\kappa^4m_Am_B\over 1024\pi^2}\left[{41\over 5}L-6S(m_A+m_B)+4\pi i{m_Am_Bm_r\over p_0t}L+\ldots\right]
\end{equation}

As in the electromagnetic case, the imaginary piece is associated with the gravitational scattering phase, or equivalently the Born iteration, which must be subtracted in order to generate a properly defined second order potential
\begin{eqnarray}
B_2^{grav}(q)&=&i\int{d^3\ell\over (2\pi)^3}{{1\over 8}\kappa^2m_A^2\over |\boldsymbol{p}_f-\boldsymbol{\ell}|^2+\lambda^2}
{i\over {p_0^2\over 2m_r}-{\ell^2\over 2m_r}+i\epsilon}{{1\over 8}\kappa^2m_B^2\over |\boldsymbol{\ell}-\boldsymbol{p}_i|^2+\lambda^2}\nonumber\\
&=&-i{\kappa^4\over 256\pi}m_A^2m_B^2{m_r\over p_0t}L
\end{eqnarray}
The result is the well-defined second order gravitational potential
\begin{eqnarray}
V_2^{grav}(r)&=&-\int{d^3q\over (2\pi)^3}e^{-i\boldsymbol{q}\cdot\boldsymbol{r}}\left({\rm Amp}_2^{grav}(\boldsymbol{q})-B_2^{grav}(\boldsymbol{q})\right)\nonumber\\
&=&-{3G^2m_Am_B(m_A+m_B)\over r^2}-{41G^2m_Am_B\hbar\over 10\pi r^3}\,,
\end{eqnarray}
which agrees with the result calculated via Feynman diagram methods by Khriplovich and Kirilin and Bjerrum-Bohr et al. in ~\cite{Khr03},\cite{Bje03}.

We can also deal with a polarizable gravitational system by use of the effective Hamiltonian
\begin{equation}
H=-{1\over 2}\alpha_G^A\sum_{ij}R^2_{0i;0j}=-{\alpha_G^A\over 2m_A^4}p_1^\alpha p_1^\gamma R_{\alpha\beta;\gamma\delta}R^{\rho\beta;\sigma\delta}p_{1\rho}p_{1\sigma}
\end{equation}
where $\alpha_G$ is the quadrupole polarizability and $R_{\alpha\beta;\gamma\delta}$ is the Riemann curvature tensor, which leads to the contact helicity amplitude for emission of a graviton pair
\begin{equation}
S^{ab}_A= t^2{\cal U}^{ij;kl}\epsilon^{a*}_{1i}\epsilon^{a*}_{1k}\epsilon^{b*}_{2j}\epsilon^{b*}_{2\ell}
\end{equation}
with
\begin{equation}
{\cal U}_A^{ij;k\ell}={\alpha_G^A\over 64}d_A^2{\cal O}_A^{ij}{\cal O}_A^{k\ell}
\end{equation}
For the case of a point mass $m_B$ interacting with a polarizable mass $m_A$, we have then
\begin{eqnarray}
&&{\rm Disc}\,{\rm Amp}_2^{N-grav}\,=-{i\over 2!}\int {d^3k_1\over (2\pi)^32k_1^0}{d^3k_2\over (2\pi)^32k_2^0}(2\pi)^4\delta^4(p_1+p_2-k_1-k_2)\nonumber\\
&\times&\sum_{a,b=1}^2{t^2\kappa^2m_B^2\xi_B^2d_B\over 8m_B}{\cal U}_A^{ij;k\ell}\epsilon^{a*}_{1i}\epsilon^{a*}_{1k}\epsilon^{b*}_{2j}\epsilon^{b*}_{2\ell}\epsilon^a_{1u}\epsilon^a_{1v}\epsilon^b_{2r}\epsilon^b_{2s}{\cal O}_B^{ur*}{\cal O}_B^{vs*}\nonumber\\
&=&{-i\kappa^2m_B\xi_B^2t^2\over 128}
\sum_{i,j,k,\ell=1}^3\sum_{u,v,r,s=1}^3<\left[{\cal U}_A^{ij;k\ell}P^T_{ik;uv}P^T_{j\ell;rs}{\cal O}_B^{ur*}{\cal O}_B^{vs*}\right]d_B>\nonumber\\
&\stackrel{y\rightarrow 1}{\longrightarrow}&{\kappa^2m_B\alpha_G^At^2\over 4096\pi}\left[J^B_{44}+6J^B_{42}+6J^B_{24}\right.\nonumber\\
&-&\left.16J^B_{33}-16J^B_{13}-16J^B_{31}+J^B_{40}+J^B_{04}+36J^B_{22}+6J^B_{20}+6J^B_{02}-16J^B_{11}+2J^B_{00}\right]\nonumber\\
&=&{\kappa^2m_B\alpha_G^At^2\over 4096\pi}\left[{2\pi m_B\over \sqrt{t}}-{163\over 35}\right]
\end{eqnarray}
so
\begin{equation}
{\rm Amp}_2^{N-grav}=-{Gm_B\alpha_G^A\over 256\pi}\left(2m_BSt^2-{163\over 35}t^2L\right)
\end{equation}
Setting $t=q^2$ and taking the Fourier transform, we find the effective potential
\begin{equation}
V_2^{N-grav}(r)=-\int{d^3q\over (2\pi)^3}e^{-i\boldsymbol{q}\cdot\boldsymbol{r}}{\rm Amp}_2^{N-grav}(\boldsymbol{q})=-{3Gm_B^2\alpha_G^A\over 32\pi}\left({1\over r^6}+{163\over 7}{\hbar\over \pi m_Br^7}\right)\label{eq:fx}
\end{equation}
which is a new result.

Finally, in the case of the long-range interaction of a pair of polarizable systems, we have
\begin{eqnarray}
&&{\rm Disc}\,{\rm Amp}_2^{NN-grav}(q)\,=-{i\over 2!}\int {d^3k_1\over (2\pi)^32k_1^0}{d^3k_2\over (2\pi)^32k_2^0}(2\pi)^4\delta^4(p_1+p_2-k_1-k_2)\nonumber\\
&\times&\sum_{a,b=1}^2{t^4}{\cal U}_A^{ij;k\ell}\epsilon^{a*}_{1i}\epsilon^{a*}_{1k}\epsilon^{b*}_{2j}\epsilon^{b*}_{2\ell}\epsilon^a_{1u}\epsilon^a_{1v}\epsilon^b_{2r}\epsilon^b_{2s}{\cal U}_B^{uv;rs}\nonumber\\
&=&{t^4\over 16}\sum_{i,j,k,\ell=1}^3\sum_{u,v,r,s=1}^3<{\cal U}_A^{ik;uv}P^T_{ik;uv}P^T_{j\ell;rs}{\cal U}_B^{uv;rs}>\nonumber\\
&=&-i{\alpha_G^A\alpha_G^Bt^4\over 32768\pi}\left[K_{44}+6K_{42}+6K_{24}-16K_{33}-16K_{13}-16K_{31}+K_{40}+K_{04}\right.\nonumber\\
&+&\left.36K_{22}+6K_{20}+6K_{02}-16K_{11}+2K_{00}\right]=-i{\alpha_G^A\alpha_G^Bt^4\over 32768\pi}{443\over 315}
\end{eqnarray}
so
\begin{equation}
{\rm Amp}_2^{NN-grav}(q)\,={\alpha_G^A\alpha_G^Bt^4\over 32768\pi^2}{443\over 630}L
\end{equation}
Taking the Fourier transform, we find the effective potential
\begin{equation}
V_2^{NN-grav}(r)=-\int{d^3q\over (2\pi)^3}e^{-i\boldsymbol{q}\cdot\boldsymbol{r}}{\rm Amp}_2^{NN-grav}(q)=-{3987\over 1024}{\alpha_G^A\alpha_G^B\over \pi^3r^{11}}
\end{equation}
which agrees precisely with the retarded form given in \cite{For15},\cite{Wu16}, when we take into account the difference in the definition of polarizability used in our paper ($\alpha_G$) and theirs $(\alpha_{1S})$---
\begin{equation}
\alpha_G^A\equiv 16\pi G\alpha_{1S}\quad{\rm and}\quad \alpha_G^B\equiv  16\pi G\alpha_{2S}
\end{equation}

These methods are also easily adapted to the case that one of the scattering particles is massless, as considered in a recent paper on the bending of light as it passes the rim of the sun~\cite{Bje15}.  An important difference is that we must use $m_A^2\xi_A^2\rightarrow -{t\over 4}$,\,\,$\tau_A\rightarrow i$ and $y\rightarrow i|y|$ with $|y|={s+{t\over 2}-m_B^2\over m_B\sqrt{t}}$.  Writing $s-m_B^2=2m_BE$, where $E$ is the incident energy of the massless particle in the laboratory frame, we work in the small angle scattering approximation $E>>\sqrt{t}$ so that $|y|>>1$, in which case, using appropriately modified values of $I_{nm}\rightarrow I_{nm}^{(0)}$,\footnote{Note that here we have divided by the normalizing factor $4Em_B$.}
\begin{eqnarray}
{\rm Disc}\,{\rm Amp}^{0-grav}_2(q)&\simeq &-i{1\over 64\pi}\kappa^4tm_B\left[4{E^3\over t^2}I^{(0)}_{00}+i8{E^2\over t\sqrt{t}}I^{(0)}_{11}-3{E\over t}J^{A(0)}_{00}\right.\nonumber\\
&-&\left.{15\over 16}{E\over t}J^B_{00}-{3\over 80}{E\over t}\right]\nonumber\\
&=&-i{\kappa^4\over 256\pi m_BE}(2m_BE+{t\over 2})^2\nonumber\\
&\times&\left[{(2m_BE+{t\over 2})^2\over t}\left({m_B\over 2E}\ln\left({2E\over m_B}\right)-{m_B^2\over 2m_BE+t}\ln\left({-m_B^2\over 2m_BE+t}\right)\right)\right.\nonumber\\
&-&\left.2(2m_BE+{t\over 2})\left({m_B\over 2E}\ln\left({2E\over m_B}\right)+{m_B\over 2m_BE+t}\ln\left({-m_B^2\over 2m_BE+t}\right)\right)\right.\nonumber\\
&+&\left.{3\over 2}L-{15\over 16}\left({\pi m_B\over \sqrt{t}}-1\right)-{3\over 80}\right]
\end{eqnarray}
so
\begin{eqnarray}
{\rm Amp}_2^{0-grav}(q)&\simeq&-{\kappa^4m_BE\over 512\pi^2}\left[{L\over t}\left[2m_BE\ln\left({2E\over m_B}\right)-(2m_BE+t)\ln\left({-{2m_BE+t\over m_B^2}}\right)\right]\right.\nonumber\\
&+&\left.3L^2+{15\over 4}\left(L+Sm_B\right)-{3\over 20}L\right]\label{eq:kf}
\end{eqnarray}
As $t\rightarrow 0$ the sum of the two terms in the top line of Eq. (\ref{eq:kf}) becomes imaginary, corresponding to a scattering phase,
\begin{equation}
B_2^{0-grav}(q)=-i{\kappa^4m_BEL\over 512\pi t}
\end{equation}
which must, as previously, be subtracted.\footnote{Using this condition, we also find that the BCJ relation is satisfied, which serves as an additional check on the calculation~\cite{Ber08}.}  The resulting potential, written in terms of the laboratory frame energy of the massless particle $E$, is
\begin{eqnarray}
V_2^{0-grav}(r)&=&-\int{d^3q\over (2\pi)^3}e^{-i\boldsymbol{q}\cdot\boldsymbol{r}}\left({\rm Amp}_2^{0-grav}(q)-B_2^{0-grav}(q)\right)\nonumber\\
&=&{15\over 4}{G^2m_B^2E\over r^2}-({15-{3\over 5})\over 4\pi}{Gm_BE\hbar\over r^3}-{12G^2m_BE\hbar\over \pi r^3}\ln{r\over r_0}
\end{eqnarray}
and agrees with the form calculated in \cite{Bje15}.

\section{Conclusion}

We have seen above how on-shell techniques can be used in order to treat higher order contributions in an entire range of spinless scattering reactions, including the electromagnetic scattering of charged systems, of charged and neutral (polarizable) systems, and of two charged systems.  Similarly in the case of gravitational interactions the interactions of two masses, of a mass and a polarizable system, and two polarizable systems can be dealt with. In each case the calculation is found to agree with previously found forms but is accomplished via an algebraic on-shell method which is considerably simpler to use than the corresponding Feynman diagram procedure.  This simplification arises due to the reordering of the Feynman integration and diagram summations and, in the gravitational case, to the use of factorization, which means that the required helicity amplitudes are given simply in terms of the product of electromagnetic amplitudes.  An additional bonus is the feature that, since we are on-shell, there is no need to include ghost contributions.  We also found that this method could be adapted to the case that one of the scattering systems becomes massless.  The result is a highly efficient method to evaluate higher order electromagnetic and gravitational scattering amplitudes.
(Note that similar methods have been used by Bjerrum-Bohr et al.~\cite{Bje14},\cite{Bje16}. The basic difference between their work and that described above is that these authors use covariant evaluation and then expand to yield the low energy forms, which requires somewhat more work than the technique described above.) Work is underway to extend our results to the case that both scattering systems are massless and to the situation that one or both of the scattering systems carries spin.

\begin{center}
{\bf\large Acknowledgement}
\end{center}

Thanks to John Donoghue for numerous discussions.  This work is supported in part by the National Science Foundation under award NSF PHY11-25915.

\end{document}